\documentclass{amsproc}

\theoremstyle{definition}

\theoremstyle{remark}

\numberwithin{equation}{section}

\begin{document}

\title{Questions on Quantization}

\author{Marc A. Rieffel}
\address{Department of Mathematics, University of California, Berkeley, CA 
94720-3840}
\email{rieffel@math.berkeley.edu}
\thanks{Supported in part by NSF grant DMS 96-13833.}

\subjclass{Primary 46L60; secondary 46L87, 81S99}
\date{February 19, 1998}

\begin{abstract}
A number of basic questions concerning quantization
within the setting of operator algebras are stated, and in the process a brief 
survey
of some of the recent developments is given.
\end{abstract}

\maketitle


I collect here a number of basic questions concerning quantization
within the setting of operator algebras, while at the same time surveying
briefly some of the recent developments. For a less up-to-date but more
leisurely survey see \cite{R8}. Here I will only give reference to relatively
recent papers. In particular, I will not repeat many important references which
already appear in \cite{R8}. An extensive treatment of many of the topics
mentioned here, including historical notes and references, will appear in a
forthcoming book by Landsman \cite{La3}. 
See \cite{Wr} for a recent comparison of various
approaches to
quantization, going beyond the operator algebra setting. 
For a fine survey of the
approach using coherent states (which has significant interaction
with operator algebras) see \cite{AG}. For an interesting collection of
questions about quantum groups in the operator algebra setting see \cite{Wa}.

The present manuscript is based on informal notes (based on earlier talks
which I gave) which I distributed
at the Dartmouth Workshop on E-theory,
Quantization, and Deformations, September 1997. A number of people 
subsequently gave me suggestions for improvements, which are incorporated
here. I am especially grateful to N. Landsman and A. Weinstein for their
suggestions.
I would, of course, very much like to
be told about any progress on answering any of the questions I ask here.

The kinematics of a classical mechanical system is given by its phase
space, which is a
manifold, say $M$, equipped with a Poisson bracket. It should be the
``shadow'' of a
quantum mechanical system, whose kinematics is given by specifying a
non-commutative
$C^*$-algebra of operators and suitable ``physical'' states on it. There
should be a
length-scale, $\hbar$, Planck's constant. As $\hbar \rightarrow 0$ the
quantum system
should converge in some sense to the classical system. We will not deal
here with dynamics.

One natural approach to finding quantum systems of which a classical system
is the
``shadow'' is to deform the pointwise product on the algebra of functions
on $M$ into a
family, parametrized by $\hbar$, of non-commutative products. There is an
extensive
literature dealing with the purely algebraic side of this, in which the
product is a
formal power-series in $\hbar$ whose coefficients are functions (and thus
the product of two
functions is not again a function). See \cite{W1} and the references
therein. But
formal power-series do not mesh well with operator algebras, and so we will
not discuss
this further here.

We look for deformed products which yield actual functions. We will work with
bounded operators, so we want bounded functions. The algebra $C_{\infty}(M)$ of
continuous complex-valued functions on $M$ which vanish at infinity
determines $M$ as
its maximal ideal space. The Poisson bracket is only defined on functions
which are
differentiable in suitable directions, so we work with dense subalgebras of
$C_{\infty}(M)$. We want $C^*$-algebras, so we must allow also deformation
of the involution
(complex conjugation) and of the $C^*$-norm (the supremum norm) on
$C_{\infty}(M)$.

\medskip \noindent
{\bf 1. Definition} \cite{R5, R8}. Let $M$ be a Poisson manifold. By a {\em
strict
deformation quantization} of $M$ we mean a dense $*$-subalgebra $A$ of
$C_{\infty}(M)$
which is carried into itself by the Poisson bracket, together with a closed
subset $I$ of the
real line
containing $0$ as a non-isolated point, 
and for each $\hbar \in I$ a product $\times_{\hbar}$, an
involution
${}^{*_{\hbar}}$, and a $C^*$-norm $\|\ \|_{\hbar}$ on the linear space of
$A$, such
that for $\hbar = 0$ they are the product, involution, and norm from
$C_{\infty}(M)$,
and such that

\begin{enumerate}
\item The completions of $A$ for the various $C^*$-norms form a continuous
field of
$C^*$-algebras over $I$.

\item For $f,g \in A$ we have
\[
\|(f \times_{\hbar} g - g \times_{\hbar} f)/\hbar - i\{f,g\}\|_{\hbar}
\rightarrow 0
\]
as $\hbar \rightarrow 0$.
\end{enumerate}
\medskip

We call attention to a further aspect which has not received much emphasis
previously.

\medskip \noindent
{\bf 2. Definition}. We will say that a strict deformation quantization is {\em
flabby} if $A$, as above, contains $C_c^{\infty}(M)$, the algebra of smooth
functions of
compact support on $M$.
\medskip

The most important present source of flabby strict deformation
quantizations is the
Weyl--Moyal quantization and its generalizations. One fairly far-reaching
generalization goes as follows \cite{R5}. Let $\alpha$ be an action of $V =
\mathbb{R}^n$ on a locally compact space $M$, and so on $C_{\infty}(M)$.
Let $A$ be
the dense
$*$-subalgebra of smooth vectors in $C_{\infty}(M)$ for $\alpha$. Let $X_j$
denote the derivation of $A$ in the $j$-th direction of $\mathbb{R}^n$ via
$\alpha$. The choice of a skew-symmetric $n \times n$ matrix $J$ determines
a Poisson bracket on $A$ defined by
$$\{f,g\}=\sum J_{jk}(X_j f)(X_k g).$$
For each $\hbar \in \mathbb{R}$ define a new product, $\times_{\hbar}$, on
$A$ by
\[
f \times_{\hbar} g = \int_{V \times V} \alpha_{\hbar
Ju}(f)\alpha_v(g)e^{2\pi iu\cdot v} dudv,
\]
where this integral must be interpreted as an oscillatory integral
\cite{R5}. Let the
involution remain undeformed. There is a natural way to define $C^*$-norms
$\|\ \|_{\hbar}$, discussed in \cite{R5}. This provides a flabby strict
deformation
quantization of $M$ with the given Poisson bracket.

Notice that the above is a {\bf universal} deformation formula, in the
sense that it works any time $\mathbb{R}^n$ acts on {\em any} space or
C*-algebra.

\medskip \noindent
{\bf 3. Question}. Are there universal
deformation formulas, at the analytical level, for actions of (at least
some) other Lie groups?
\medskip

At the
Lie algebra level and in terms of formal power series, a few such universal
deformation formulas are known. See \cite{GZ} and references therein. In
particular, there is a universal deformation formula for the Lie algebra of
the affine group of the real line (and a related formula appears in
slightly buried form in \cite{Oh}). This suggests that at least for actions
of the affine group there should be a universal deformation formula at the
analytical level.

For actions of $\mathbb{R}^n$ many
specific examples of quantum spaces constructed as strict deformation
quantizations are given in \cite{R5, R8}, such as quantum disks, tori and
spheres. Quantum groups can be
constructed by
this method \cite{R6, R9, Ln}. Closely related constructions provide
quantum Heisenberg
manifolds and lens spaces \cite{R1, Ab, AE1, AE2}, as well as an algebra
for the
space-time uncertainty relations \cite{R10}. (Probably the closely related
constructions in \cite{LR1, LR2} can be arranged, in the appropriate
situations, to give flabby strict deformation quantizations, though I have
not checked this.) But it is unclear to me how often the above construction
applies, because I have found little information about:

\medskip \noindent
{\bf 4. Question}. Given a Poisson bracket, how does one determine whether
it comes from an action of $\mathbb{R}^n$ as described above, for some $n$?
In particular, what cohomological obstructions are there to expressing a
Poisson bracket globally in terms of a family of {\em commuting}
vector fields?
\medskip

In answer to my query following a suggestion of Alan Weinstein, 
Charles Pugh very recently showed me a proof that any action of $\mathbb{R}^n$
on the 2-sphere, for any $n$, must have a fixed point. This shows that
no Poisson bracket on the 2-sphere which comes from a symplectic
structure can be given by an action of $\mathbb{R}^n$. (But the question
of whether there exists some strict deformation quantization of the 2-sphere
for some symplectic structure remains open.) 

As a particularly nice situation in which one can ask Question 4 we have:

\medskip \noindent
{\bf 5. Question}. Let $M$ be a manifold with a Poisson bracket coming from
an action of $\mathbb{R}^n$ as above. Let $\alpha$ be a free and proper
action of a group $G$ on $M$ which preserves the Poisson bracket, so that
one obtains a Poisson bracket on the quotient manifold $M/\alpha$. When
does this latter Poisson bracket come from an action of $\mathbb{R}^n$ on
$M/\alpha$? Also, the same question except for $\alpha$ replaced by a more
general suitable equivalence relation on $M$?
\medskip

A particular case of the process just described occurs in the construction
of the quantum Heisenberg manifolds discussed in \cite{R1, Ab, AE1, AE2}.
For that case it is shown at the end of \cite{AE2} that the Poisson
bracket does in fact come from an action of $\mathbb{R}^2$. But it is
not clear to me why this should be true in general.

\medskip \noindent
{\bf 6. Question}. Given a Poisson manifold $M$, how does one determine
whether there is a Poisson manifold $N$ whose Poisson bracket comes from an
action of $\mathbb{R}^n$, and a group action $\alpha$ (or more general
equivalence relation) on $N$ preserving the Poisson bracket, such that
$N/\alpha \cong M$ as Poisson manifolds?
\medskip

The reason for asking this question is that even when the answer
to Question 5 is negative, one can hope to construct a deformation
quantization of the quotient manifold from one of the big manifold:

\medskip \noindent
{\bf 7. Question}. Given a Poisson manifold and a quantization of it, and
nice actions of a group on both the manifold and the quantization which are
compatible, how does one construct from this a quantization of the quotient
Poisson manifold?
\medskip

In the literature there is some discussion of this question for formal
deformation quantizations \cite{GRZ, X1, F}. At a heuristic level one must
form a generalized fixed-point algebra of the quantization (as done in
\cite{R7} for the special case of actions of $\mathbb{R}^n$). But the
technicalities are elusive, and certainly involve the definition of
``nice'', which probably involves a suitable notion of ``proper action'' on
non-commutative C*-algebras, as discussed in \cite{R2} for exactly these
purposes.

Another collection of examples of flabby strict deformation quantizations
arises from
nilpotent Lie algebras $\mathfrak{g}$. Let $M = \mathfrak{g}^*$ denote the
dual vector
space of $\mathfrak{g}$, with the well-known linear Poisson bracket from
$\mathfrak{g}$ defined by
$$\{f,g\}(\mu)= \langle [df(\mu), dg(\mu)], \mu \rangle.$$ For each $\hbar
\in \mathbb{R}$ define a new Lie bracket $[\ ,\ ]_{\hbar}$ on
$\mathfrak{g}$ by $[X,Y]_{\hbar} = \hbar[X,Y]$, so that we are contracting
$\mathfrak{g}$ to
an Abelian Lie algebra. Let $G_{\hbar}$ denote $\mathfrak{g}$ with the
corresponding Lie
group structure, and let $*_{\hbar}$ denote the corresponding convolution
of functions on
$\mathfrak{g}$. Let $A = {S}$ be the algebra of Schwartz functions on
$\mathfrak{g}^*$, and let
$\ ^\wedge$ and $\ ^\vee$ denote the Fourier transform from
$\mathfrak{g}^*$ to $\mathfrak{g}$
and its inverse. For $f,g \in A$ set
\[
f \times_{\hbar} g = ({\hat f} *_{\hbar} {\hat g})^{\vee}. \]
With the $C^*$-norms coming from the group $C^*$-algebras $C^*(G_{\hbar})$
this defines
a flabby strict deformation quantization of $\mathfrak{g}^*$ \cite{R3}.
This construction too can be used to construct certain non-compact quantum
groups \cite{R4, Ka1, Ka2, Ka3}.

\medskip \noindent
{\bf 8. Question}. To what extent can
more general contractions of Lie groups be seen to give strict deformation
quantizations of ``non-commutative Poisson algebras'' as defined in
\cite{BG, N1, N2, X2}?
\medskip

Brief allusion to this possibility occurs following conjecture 4.29 of
\cite{BCH}.

\medskip

But there are interesting strict deformation quantizations which are not
flabby. In the example just above, if one lets $\mathfrak{g}$ be a
non-nilpotent exponential solvable Lie group, then convolution does not
carry the
Schwartz functions into themselves, and so one must take as $A$ the smooth
functions whose
Fourier transforms have compact support \cite{R3}. Thus the elements of $A$
are analytic functions, and so this example is not flabby. In fact, many of
the non-flabby examples feel rigid in the sense that
analytic functions do. (Does this permit the use of analytic continuation
to get more
precise information?)
This brings us to the most basic questions:

\medskip \noindent
{\bf 9. Question}. When does a Poisson
manifold admit a strict deformation quantization? In particular, what
cohomological obstructions are there to having a strict deformation
quantization?
\medskip

\medskip \noindent
{\bf 10.Question}. When does a Poisson manifold admit a flabby strict
deformation
quantization. In particular, what cohomological obstructions are there? Do
there exist Poisson manifolds which admit a strict deformation quantization
but do not admit one which is flabby?
\medskip

My guess is that the answer to this last question is ``yes". But it is 
striking that up to now there are virtually no negative results for
these two questions. It would be interesting to see to what extent
the techniques in \cite{WX, X3}, which are concerned with formal deformation
quantization, can be adapted to the operator algebra setting, and whether
they are of any help in dealing with the above two questions. Very
recently Kontsevich \cite{Ko} has shown that every Poisson manifold
has a formal deformation quantization. It is quite a challange to see 
whether any of his ideas can be adapted to the operator algebra setting.

Many interesting examples of strict deformation quantizations have been
constructed in terms of generators and relations, including many quantum
groups. (Construction by generators and relations works best for compact
spaces, since otherwise the generators tend to give unbounded operators
which are technically difficult to work with, though Woronowicz has
nevertheless done remarkable things in the situation of unbounded
operators.)
But most of the examples constructed in this way are not flabby. Nagy
\cite{N1, N2} has given a framework for proving that many such
constructions give strict deformation quantizations. This framework has
been reinforced by Blanchard \cite{Bln}. It would be interesting to know if
the results of \cite{BEW} can be brought within this framework.

A more geometrical approach to constructing deformation quantizations
is through groupoid C*-algebras. See \cite{Sh4, Sh5, Sh6} and the
references therein. Again it would be interesting to know what can
be said about the flabbiness of the resulting quantizations, as
well as their relation to a number of the questions which follow.

It is important to know how a deformation quantization relates to the
geometry of the Poisson manifold. A Poisson manifold decomposes into
symplectic leaves, and it is natural to expect that these symplectic leaves
might correspond somewhat to (two-sided) ideals in the C*-algebras of the
deformation quantization. If a leaf is not closed, then we would only
expect that its closure might correspond to an ideal. (Then, in the nice
case in which the leaf is open in its closure, the leaf itself 
might correspond to an
ideal in the quotient algebra.)

\medskip \noindent
{\bf 11. Definition}. We say that a deformation quantization of a Poisson
manifold (with notation as above) is {\bf leaf-preserving}, or {\bf
tangential}, if for each
symplectic leaf $L$ the ideal
$$
I_L = \{f \in A: f|_L = 0\}
$$
of $A$ is also a *-ideal for each of the deformed products and 
involutions. We also
require that the closure $\bar L$ of $L$ is determined by $I_L$ in 
the sense that
$\bar L$
consists exactly of every point at which all the functions in $I_L$ vanish.
\medskip

The generalized Weyl-Moyal quantization for actions of $\mathbb{R}^n$
described above is easily seen to be leaf-preserving, using results in
\cite{R5}. But the flabby strict deformation
quantization described above for nilpotent Lie groups is well--known to not
be leaf-preserving most of the time. This appears to be due to the fact
that the Fourier transform does not mesh naturally with the group
structure. There is evidence that by modifying the Fourier transform to an
``adapted Fourier transform'', the deformation quantization for nilpotent
Lie groups, and even for exponential groups, can be modified so as to be
leaf preserving. See \cite{Be} and references therein. But the full extent
to which this
can be done successfully at the analytical level does not seem to have been
worked out yet. But this raises the more general:

\medskip \noindent
{\bf 12. Question}. Does every Poisson manifold which admits a flabby
strict deformation quantization, always admit one which is
leaf-preserving?
\medskip

\medskip \noindent
{\bf 13. Question}. Are there examples of Poisson manifolds which admit a
(non-flabby) strict
deformation quantization, but do not admit one which is leaf-preserving?
\medskip

My guess is that such examples exist. Within the setting of formal deformation
quantization this question has been discussed recently in \cite{C2, Ls}.
For the geometric analogue see \cite{W2}.

In many important situations in which Poisson manifolds arise there is
additional
structure present which one would like to preserve under quantization.
Probably the most common extra structure is a group of symmetries, that is,
a group of diffeomorphisms of the manifold which respect the Poisson
bracket. One can then ask most of the above questions but with the added
requirement that the
additional structure be preserved, in a suitable sense. In this setting we
have one of the very few ``no--go'' examples that I know of:

\medskip \noindent
{\bf 14. Example} (at the very end of \cite{R1}). Consider the two-sphere
with the
Poisson bracket from its rotationally invariant symplectic structure. There
is no strict deformation quantization of this Poisson manifold which
preserves the action of $SO(3)$.
\medskip

It would be interesting to know how common such examples are. That is, we
have the following counterpart to Question 9:

\medskip \noindent
{\bf 15. Question}. Given a Poisson manifold with action of a group, what
are some necessary conditions for the existence of a strict deformation
quantization which respects the group action? In particular, what
cohomological obstructions are there to such existence?
\medskip

By means of generators and relations
Nagy \cite{N1} has constructed a strict deformation quantization of the
disk which respects the action of $SL(1,1)$. This example is not flabby.
But the disk with its $SL(1,1)$-invariant Poisson bracket is
symplectomorphic to the plane with its standard Poisson bracket, and so
admits a strict deformation quantization by the Weyl-Moyal construction,
which is flabby, but does not
preserve the $SL(1,1)$-action.

\medskip \noindent
{\bf 16. Question}. Is there a flabby strict deformation quantization of
the disk which preserves the action of $SL(1,1)$?
\medskip

I would not be surprised if the answer is ``no''.

A related type of additional structure which can be present is the
canonical coproduct on the algebra of functions on a Lie group, which one
wants to preserve (perhaps in a deformed way) when one is trying to deform
the Lie group into a quantum group. For the standard quantum group
$SU_{\mu}(n)$ of Woronowicz (constructed by generators and relations), it
was shown by
Sheu \cite{Sh1} for $n = 2$, and then by Nagy \cite{N1, N2} for general
$n$, that they form a strict deformation quantization of $SU(n)$. But it is
not flabby. Related results about deformations and quantum groups can be
found in \cite{Bln}.

In this context we have the only other substantial ``no-go'' example of
which I am aware:

\medskip \noindent
{\bf 17 Example}. Sheu \cite{Sh2, Sh3} has shown that there is no strict
deformation quantization of $SU(n)$ with its standard Poisson-Lie bracket
which simultaneously preserves the comultiplication and is leaf-preserving.
\medskip

Again, it would be desirable to know how common such examples are. But note
in contrast that the strict deformation quantizations of, say, $SU(n)$ for
$n\geq 3$
as quantum groups which are constructed in \cite{R6} for 
non-standard Poisson-Lie
brackets, are leaf-preserving (and flabby) and preserve the
comultiplication.

Given a strict deformation quantization, it is natural to ask whether the
deformed C*-algebras have the same ``algebraic topology" as the original
manifold. In particular, do they have isomorphic K-groups? (In discussing
this one probably wants to assume that the set $I$ over which
$\hbar$ ranges is connected.) Already one sees
from the quantum tori that the order structure on $K_0$ will often be
different.
For the generalized Weyl-Moyal quantization discussed above it is shown in
\cite{R7} that
the K-groups are isomorphic. Very
recently Nagy
\cite{N3} has developed techniques in $E$-theory which deal with the case
in which many
fiber algebras are non-isomorphic. He applies this to show that the
K-groups are isomorphic in many other cases, in particular for certain
quantum groups.

However, K-groups are not always preserved under strict deformation
quantization. For example, let $M$ be the closed unit interval, with
the $0$ Poisson bracket on $C(M)$. (There are easy variations on this
example in which $M$ is a compact manifold without boundary.) Let $A$
be the dense subalgebra consisting of the polynomials. Keep the product
and involution fixed, and deform only the C*-norm on $A$, as follows.
Let $N_\hbar$ denote the union of the interval $[\hbar, 1]$ with the
sequence ${1/n: n\geq 1}$. Let $\|\ \|_\hbar$ be the supremum norm
over $N_\hbar$. Since $N_\hbar$ has an infinite number of components, the
completion of $A$ for each of these norms will have infinitely-generated
$K_0$  group, in contrast to $C(M)$. In this example one can also take
as $A$ the functions which are analytic in a neighborhood of the
interval $[0,1]$, so as to obtain a ``local C*-algebra'' as defined
in \cite{Bla}, where the comment in 3.1.6 of \cite{Bla} about partitions
of unity is relevant. However, this example is not flabby, and
it is possible that flabbiness is useful in connection with the K-theory of
quantization. That is:

\medskip \noindent
{\bf 18. Question}. Are the K-groups of the C*-algebra completions
of the algebras of any flabby strict deformation quantization all
isomorphic?
\medskip

We remark that the quantum group examples treated by Nagy mentioned above
are not flabby but nevertheless the K-groups are isomorphic. We also remark
that the $0$ Poisson bracket can have non-commutative strict deformation
quantizations. If in any strict deformation quantization of a non-zero
Poisson bracket one reparametrizes by replacing $\hbar$ by $\hbar^2$
one obtains a strict deformation of the $0$ Poisson bracket.

A Poisson manifold has a smooth structure, with its associated deRham
cohomology. It would be desirable to have this cohomology preserved under
strict deformation quantization. In the non-commutative case the role of
the deRham cohomology is played by Connes' cyclic homology. This requires
the use of a dense *-subalgebra to play the role of the algebra of smooth
functions. But in many of the examples constructed by generators and
relations, the most evident dense *-subalgebra feels more like an algebra
of polynomials than like the algebra of all smooth functions. In
particular, its K-groups and cyclic homology are unlikely to agree with
those of the original manifold. Already for the basic example of
Woronowicz' quantum group $SU_{\mu}(2)$ the situation is unclear. That is:

\medskip \noindent
{\bf 19. Question}. For the quantum group $SU_{\mu}(2)$ (and also for
$SU_{\mu}(n))$ , and for any
given $\mu$, is there a dense *-subalgebra $A$ such that: 
\begin{enumerate}
\item The algebra $A$ is closed in $SU_{\mu}(2)$ 
under the holomorphic functional
calculus, so that it has the same K-groups 
as ($SU_{\mu}(2)$ and so as) $SU(2)$. 
\item The periodic
cyclic homology of the algebra $A$ agrees in the appropriate sense with the
deRham cohomology of $SU(2)$. 
\item The comultiplication of $SU_{\mu}(2)$
carries $A$ into some kind of smooth tensor product $A\otimes A$.
\end{enumerate}
\medskip

It seems to me that only with an affirmative answer can one really say that
$SU_{\mu}(2)$ has the structure of a non-commutative differentiable
manifold with a smooth ``group structure'', so that it is not just a
quantum group, but in fact a quantum Lie group.

When $Q$ is a manifold, its cotangent bundle, $T^*Q$, carries a canonical 
symplectic form. When $Q$ is the configuration space of a classical
mechanical system, $T^*Q$ is its phase space. Thus the problem of
quantizing $T^*Q$ is one of central importance. In many physical situations
there is a natural Riemannian metric on $Q$. One then expects to need
to use this Riemannian metric to obtain a natural quantization of $T^*Q$. There
is an extensive literature on this subject. Recent discussion appears
in \cite{LQ, La1, Om}. A substantial discussion of this will appear
in \cite{La3}. But most of the constructions go in the direction of
producing a quantization in (at most) the weaker sense of Definition 23
below. Thus there is need for a clearer answer to:

\medskip \noindent
{\bf 20. Question}. How often, and in what ways, can a suitable Riemannian
metric on $Q$ be used to obtain a strict deformation quantization
of $T^*Q$?
\medskip

There are recent generalizations of the Weyl quantization to Riemannian
symmetric spaces and related situations
which appear to have much promise of giving strict deformation
quantizations \cite{U1, U2, UU}.
But no general proofs have been given yet.

Landsman (\cite{La2, LW} and references therein) has extensively discussed
how ideas from
induced representation of $C^*$-algebras provide the quantum version of the
classical
technique of symplectic reduction. He indicates in particular how this
applies to
strict deformation quantization. This provides further examples, but there
is much more remaining to be developed in this direction.

There is no uniqueness for strict deformation quantizations, as seen
already by considering quantum tori \cite{R8}. But the example of quantum tori
suggests that there should be some natural definition of equivalence
of strict deformation quantizations which is weak enough to permit
the fiber algebras to be non-isomorphic. Thus:

\medskip \noindent
{\bf 21 Question}. What are some useful definitions of equivalence
for strict deformation quantization?
\medskip

Even in the absence of uniqueness there are many reasons
to believe that various kinds of additional structure on a Poisson manifold
may select a natural quantization. At the level of formal deformation
quantization of symplectic manifolds this has been realized through the use
of symplectic connections by Fedosov. (See the review in \cite{W1}.) 
Speculation that the presence of a Riemannian metric on the Poisson
manifold (not just on a configuration space) can serve the same purpose
is given in \cite{Kl, KM}. So it is natural to ask:

\medskip \noindent
{\bf 22. Question}. What types of additional structure on a Poisson
manifold will lead to a canonical choice of a strict deformation
quantization? (So in particular, to what extent will the specification of a
Riemannian metric do this?)
\medskip

For many situations it is probably unreasonable to ask for a strict
deformation quantization.
There is a weakening of the quantization requirements which has permitted
many more
examples to be constructed. This is often associated with the names Berezin
and Toeplitz. But as suggested by Landsman, it is probably best to refer to
it simply as ``strict quantization'', reserving the names Berezin and
Toeplitz for more specific versions of it. We now no longer ask for a deformed
product on a Poisson algebra $A$ of
functions.

\medskip \noindent
{\bf 23. Definition}. Let $M$ be a Poisson manifold. By a {\em strict
quantization} of $M$ we mean a dense $*$-subalgebra $A$ of $C_{\infty}(M)$
which is carried into itself by the Poisson bracket, together with a closed
subset $I$ of the real line containing $0$ as a non-isolated point, 
and for each $\hbar \in I$ a
linear map $T_{\hbar}$ (usually preserving the
involution) of $A$ into a C*-algebra $\bar A_{\hbar}$ which is generated by
the range of $T_{\hbar}$, such
that for $\hbar = 0$ the map $T_0$ is the canonical inclusion of $A$ into
$C_{\infty}(M)$, and such that

\begin{enumerate}
\item The maps $\hbar \rightarrow T_{\hbar}(a)$ define the structure of a
continuous field of C*-algebras on the family $\{\bar A_{\hbar}\}$.

\item For $f, g \in A$ we have
\[
\|(T_{\hbar}(f)T_{\hbar}(g) - T_{\hbar}(g)T_{\hbar}(f))/\hbar -
iT_{\hbar}(\{f,g\})\|_{\hbar} \rightarrow 0 \]
as $\hbar \rightarrow 0$.
\end{enumerate}

If each $T_{\hbar}$ is injective, we will say that the strict quantization
is ``faithful''.
\medskip

The difference from deformation quantization is that for $f,g \in A$ the
product operator $T_{\hbar}(f)T_{\hbar}(g)$
need not be of the form $T_{\hbar}(h)$ for some $h \in A$, so no product is
defined on
$A$.

Most of the above questions have their counterparts in this weaker setting.
We will not spell them out here. However, the precise relation with strict
deformation quantization is unclear to me. In particular:

\medskip \noindent
{\bf 24. Question}. Is there an example of a Poisson manifold which admits
a (faithful?) strict quantization, but does not admit a strict deformation
quantization?
\medskip

More specifically:

\medskip \noindent
{\bf 25. Question}. Is there an example of a faithful strict quantization,
$T$, defined on a dense *-subalgebra $A$, such that it is impossible to
find a dense *-subalgebra $B \subseteq A$ on which $T$ is a strict {\em
deformation} quantization, that is, such that for any $f, g \in B$ and any
$\hbar$ the product operator $T_{\hbar}(f)T_{\hbar}(g)$ is of the form
$T_{\hbar}(h)$ for some $h \in B$, where $h$ depends on $\hbar$?
\medskip

When $T_{\hbar}$ is injective, it is appropriate to consider $\bar
A_{\hbar}$ to be a quantum version of the Poisson manifold $M$. It then becomes
very interesting to determine the properties of this $C^*$-algebra (e.g.
its structure and
its algebraic topology). A certain amount is known in some cases, as referenced
in \cite{R8}. Further information and some interesting
conjectures about the structure of other examples can be found in
\cite{BLR} . It would be interesting to know if the results of \cite{C1}
and its predecessors could also be brought within this setting.

In many of the examples which have been constructed, the $T_{\hbar}$'s are
suitable
Toeplitz operators, that is, they are compressions to suitable subspaces of
a representation of $C_{\infty}(M)$. See references in \cite{R8, AG}, and
the more recent papers
\cite{BMS, Bo, Sc} and their references. We will refer to such a
quantization as a ``strict Toeplitz quantization''. In this case the
quantization is clearly ``positive'' in the sense that positive functions
are taken to positive operators.
But for the Weyl-Moyal-type quantizations positivity usually fails.

\medskip \noindent
{\bf 26. Question}. Is it impossible for a strict deformation quantization
to be positive?

\medskip \noindent
{\bf 27. Question}. If a Poisson manifold admits a strict quantization,
does it always admit one which is positive?
\medskip

For many of the examples which have been constructed,
especially for compact $M$, the C*-algebras $\bar A_{\hbar}$ are
finite-dimensional (see \cite{BMS, Bo, Sc} and references therein), so that
it does not seem
appropriate to consider them to
be quantum versions  of $M$. Thus it would be very interesting to have
alternative
constructions which produce $T_{\hbar}$'s which are injective.

\medskip \noindent
{\bf 28. Question}. If a Poisson manifold admits a strict quantization,
does it always admit one which is faithful?
\medskip

There is one important situation in which recently this has been
successfully shown to be true.
There has been much interest in the construction of quantum versions of
compact Riemann surfaces of genus $\ge 2$. Most of the attempted
constructions run into this
problem of producing finite dimensional Hilbert spaces. However, Klimek and
Lesniewski
\cite{KL} found a way around this difficulty by using a cleverly-chosen
non-compact
covering of the Riemann surface with covering group $\mathbb{Z}$,
constructing a
Toeplitz quantization of this covering surface respecting the action of
$Z$, and showing
that there are plenty of $\mathbb{Z}$-invariant operators, which then give the
quantization
of the compact surface. The exploration of the properties of these quantum
Riemann
surfaces should be an interesting adventure in the years ahead. (I have
learned recently that T. Natsume and R. Nest have developed a
somewhat different approach to constructing quantum Riemann surfaces. It
starts with the action of the fundamental group of the Riemann surface on
the quantization of the disk by Toeplitz operators. They then form the
cross product algebra, and then cut down by carefully chosen projections.
They are able to obtain substantial information about the properties of the
resulting algebras.)

\bibliographystyle{amsalpha}

\end{document}